\documentclass[aps,showpacs,floatfix,twocolumn,byrevtex,superscriptaddress,longbibliography]{revtex4-1}

\usepackage{amsmath}
\usepackage{amssymb}
\usepackage{amstext}
\usepackage{amsopn}
\usepackage{amsfonts}
\usepackage{amsxtra}
\usepackage[english]{babel}
\usepackage{amsmath}
\usepackage{graphicx}
\usepackage{float}
\usepackage{bm}
\usepackage{multirow}
\usepackage{dcolumn}
\usepackage{hyperref}
\usepackage{CJK}

\begin{document}

\title{Unravelling the mechanism of the semiconducting-like behavior and its relation to superconductivity
in (CaFe$_\mathbf{1-x}$Pt$_\mathbf{x}$As)$_\mathbf{10}$Pt$_\mathbf{3}$As$_\mathbf{8}$}
\author{Run Yang}
\affiliation{Beijing National Laboratory for Condensed Matter Physics, Institute of Physics,
  Chinese Academy of Sciences, Beijing 100190, China}
\affiliation{Condensed Matter Physics and Materials Science Division, Brookhaven National Laboratory,
  Upton, New York 11973, USA}
\affiliation{School of Physical Sciences, University of Chinese Academy of Sciences, Beijing 100049, China}
\author{Yaomin Dai}
\affiliation{Center for Superconducting Physics and Materials, National Laboratory of
  Solid State Microstructures and Department of Physics, Nanjing University, Nanjing 210093, China}
\author{Jia Yu}
\author{Qiangtao Sui}
\author{Yongqing Cai}
\affiliation{Beijing National Laboratory for Condensed Matter Physics, Institute of Physics,
  Chinese Academy of Sciences, Beijing 100190, China}
\affiliation{School of Physical Sciences, University of Chinese Academy of Sciences, Beijing 100049, China}
\author{Zhian Ren}
\affiliation{Beijing National Laboratory for Condensed Matter Physics, Institute of Physics,
  Chinese Academy of Sciences, Beijing 100190, China}
\affiliation{School of Physical Sciences, University of Chinese Academy of Sciences, Beijing 100049, China}
\affiliation{Collaborative Innovation Center of Quantum Matter, Beijing 100084, China}
\author{Jungseek Hwang}
\affiliation{Condensed Matter Physics and Materials Science Division, Brookhaven National Laboratory,
  Upton, New York 11973, USA}
\affiliation{Department of Physics, Sungkyunkwan University, Suwon, Gyeonggi-do 16419, Korea}
\author{Hong Xiao}
\affiliation{Center for High Pressure Science and Technology Advanced Research, Beijing 100094, China}
\author{Xingjiang Zhou}
\affiliation{Beijing National Laboratory for Condensed Matter Physics, Institute of Physics,
  Chinese Academy of Sciences, Beijing 100190, China}
\affiliation{School of Physical Sciences, University of Chinese Academy of Sciences, Beijing 100049, China}
\affiliation{Collaborative Innovation Center of Quantum Matter, Beijing 100084, China}
\author{Xianggang Qiu}
\email[]{xgqiu@iphy.ac.cn}
\affiliation{Beijing National Laboratory for Condensed Matter Physics, Institute of Physics,
  Chinese Academy of Sciences, Beijing 100190, China}
\affiliation{School of Physical Sciences, University of Chinese Academy of Sciences, Beijing 100049, China}
\affiliation{Collaborative Innovation Center of Quantum Matter, Beijing 100084, China}
\author{Christopher C. Homes}
\email[]{homes@bnl.gov}
\affiliation{Condensed Matter Physics and Materials Science Division, Brookhaven National Laboratory,
  Upton, New York 11973, USA}
\date{\today; version 6}
%
%

\begin{abstract}
The temperature-dependence of the in-plane optical properties of (CaFe$_{1-x}$Pt$_{x}$As)$_{10}$Pt$_{3}$As$_{8}$
have been investigated for the undoped ($x=0$) parent compound, and the optimally-doped ($x= 0.1$) superconducting
material ($T_{c}\simeq 12$~K) over a wide frequency range.  The optical conductivity has been described using
two free-carrier (Drude) components, in combination with oscillators to describe interband transitions.
At room temperature, the parent compound may be described by a strong, broad Drude term, as well as a
narrow, weaker Drude component.  Below the structural and magnetic transitions at $\simeq 96$ and 83~K,
respectively, strength is transferred from the free-carrier components into a bound excitation at
$\simeq 1000$~cm$^{-1}$, and the material exhibits semiconducting-like behavior.
In the optimally-doped sample, at room temperature the optical properties are again described by narrow
and broad Drude responses comparable to the parent compound; however, below $T^\ast \simeq 100$~K,
strength from the narrow Drude is transferred into a newly-emergent low-energy peak at $\simeq 120$~cm$^{-1}$,
which arises from a localization process, resulting in semiconducting-like behavior.
Interestingly, below $T_{c}$, this peak also contributes to the superfluid weight,
indicating that some localized electrons condense into Cooper pairs; this observation may provide
insight into the pairing mechanism in iron-based superconductors.
\end{abstract}


\maketitle

%
%
\section{Introduction}
The discovery of iron-based superconductors has prompted in an intensive
investigation of this class of materials in the hope of discovering new
compounds with high superconducting critical temperatures ($T_c$'s)
\cite{Johnston2010,Paglione2010,Si2016}.
In both iron-based superconductors (FeSCs) and cuprates, a variety of unusual
normal-state phenomena are observed that are believed to have an important connection to the
superconductivity (SC) \cite{Lee2006,Wang2016}.  In the optimally-doped cuprates the
resistivity often shows a peculiar non-saturating linear temperature dependence
that at high temperature may violate the Mott-Ioffe-Regel limit \cite{Hussey04},
leading it to be described as a marginal Fermi liquid \cite{Varma89}.
A pseudogap develops in underdoped regime well above the critical temperature
($T_{c}$) \cite{Timusk1999}, which has been interpreted as evidence for preformed
Cooper pairs without global phase coherence~\cite{Yang2008,Wang2005,Xu2000};
on the other hand, competing orders, such as charge-ordered states, have also
been proposed as the origin of this feature \cite{Valla2006}.
%
%
In FeSCs, one of the most interesting phenomena in the normal state is the emergence
of nematicity, or rotational symmetry breaking of the electronic states \cite{Fernandes2012};
however, its origin and relation to the superconductivity in these materials is still
uncertain \cite{Wang2016}.

\begin{figure}[tb]
\includegraphics[width=2.7in]{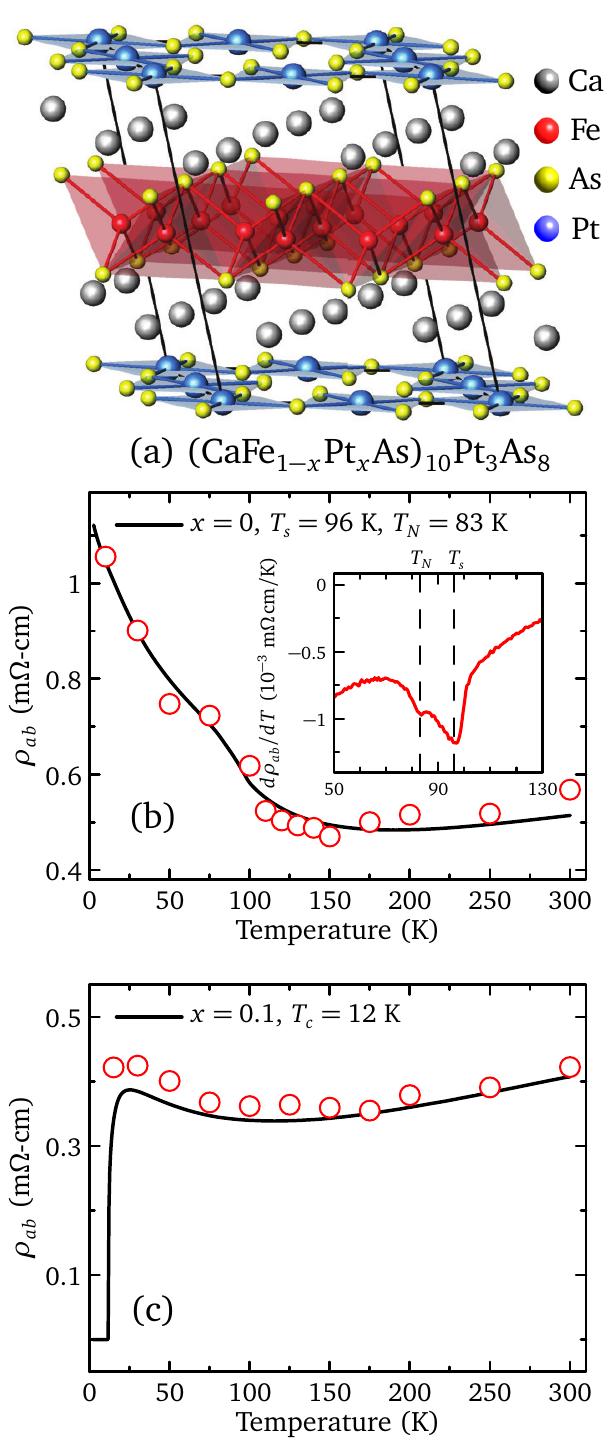}
\caption{(a) The triclinic unit cell of (CaFe$_{1-x}$Pt$_{x}$As)$_{10}\-$Pt$_{3}$As$_{8}$
showing the Fe--As layers separated by Ca and Pt$_3$As$_8$ sheets.
(b) The in-plane resistivity of the $x=0$ (undoped) sample showing a semiconducting-like
response at low temperature.  Inset: $d\rho_{ab}/dT$ showing local minima at $T_s$ and
$T_N$. (c) The resistivity for the $x=0.1$ (optimally-doped) sample, again showing
a semiconducting response at low temperature just above $T_c$.  The circles denoting
$\sigma_{1}(\omega\rightarrow 0) \equiv \sigma_{dc}$ are in good agreement with the
transport data.}
\label{fig:resis}
\end{figure}

%
%
The (CaFe$_{1-x}$Pt$_{x}$As)$_{10}$Pt$_{3}$As$_{8}$ (Ca 10-3-8) materials
exhibit some rather interesting properties.  The unit cell of the undoped parent compound
is shown in Fig.~\ref{fig:resis}(a); the conducting Fe--As layers are separated by Ca
atoms and insulating Pt$_3$As$_8$ layers, resulting in an inter-layer distance as large
as 10.6~{\AA}.  Transport measurements indicate this material is highly two dimensional
(2D) \cite{Ni2011}.   The phase diagram [Fig.~4(a) of Ref.~\onlinecite{Gao2014}]
indicates that the parent compound is an antiferromagnetic (AFM) semiconductor;
the resistivity and other experimental probes \cite{Cho2012,Zhou2013,Sturzer2013,Sapkota2014}
indicate that this material undergoes structural and magnetic transitions at
$T_s \simeq 96$~K and $T_N \simeq 83$~K, respectively.
Through the application of pressure, or by doping Pt on the Fe site (electron doping),
the AFM order is suppressed, and superconductivity emerges with a maximum $T_c\simeq 12$~K in
the optimally-doped material \cite{Xiang2012}.  However, the semiconducting-like
behavior still remains (resistivity increases upon cooling) above the AFM and SC dome
\cite{Gao2014,Ni2011}, which is reminiscent of the pseudogap-like behavior in
cuprates~\cite{Yang2008}.  Investigating the origin of such distinct behavior and how
it evolves into a superconductor may provide insight into the pairing mechanism in
iron-based superconductors.

%
%
%

%
%
In this work the temperature dependence of the in-plane optical properties of
(CaFe$_{1-x}$Pt$_{x}$As)$_{10}$Pt$_{3}$As$_{8}$ for undoped ($x=0$) and optimally-doped
($x=0.1$) samples is investigated.  The real part of the optical conductivity is
particularly useful as it yields information about the free-carrier response and interband
transitions; in the zero-frequency limit, the dc conductivity is recovered, allowing
comparisons to be made with transport data.
%
%
The optical properties suggest that the semiconducting-like behavior in the parent
compound likely originates from AFM order that leads to a reconstruction of the Fermi
surface and a decrease in the carrier concentration.  In the optimally-doped sample,
torque magnetometry indicates superconducting fluctuations well above $T_c$, suggesting
that this material may not be homogeneous.  Along with the semiconducting-like behavior,
at low temperature we observe the emergence of a peak in the optical conductivity in the
far-infrared region, which is attributed to localization driven by either scattering from
impurities or AFM spin fluctuations.  Interestingly, below $T_{c}$, this peak also contributes
to the superfluid weight, indicating that there is likely a relationship between magnetism
and superconductivity.

%
%
\section{Experiment}
High-quality single crystals of (CaFe$_{1-x}$Pt$_{x}$As)$_{10}\-$Pt$_{3}$As$_{8}$ with
good cleavage planes (001) were synthesized using self-flux method~\cite{Ni2013}.
The temperature dependence of the in-plane resistivity for the undoped and optimally-doped
materials is shown in Figs.~\ref{fig:resis}(b) and \ref{fig:resis}(c), respectively.
At room temperature, the resistivity of both materials is comparable.  In the undoped
material, $\rho_{ab}$ exhibits relatively little temperature dependence until
$\simeq 150$~K, below which it exhibits a semiconducting response, increasing gradually,
with inflection points at $T_N\simeq 83$ and $T_s\simeq 96$~K, shown in the inset of
Fig.~\ref{fig:resis}(b).
The resistivity of the doped material initially decreases as the temperature is reduced
and then undergoes a slight upturn resulting in a broad minimum at about 100~K; below
$T_c \simeq 12\,$K the resistivity abruptly drops to zero.
The reflectance from freshly-cleaved surfaces has been measured over a wide temperature
($\sim 5$ to 300~K) and frequency range ($\sim 2$~meV to about 5~eV) at a near-normal
angle of incidence for light polarized in the \emph{a-b} planes using an \emph{in situ}
evaporation technique~\cite{Homes93}.  The complex optical properties have been determined
from a Kramers-Kronig analysis of the reflectivity.  The reflectivity is shown in Supplementary
Figs.~S1(a) and S1(b); the details of the Kramers-Kronig analysis are described in the
Supplementary Material \cite{Suplmt}.  Magnetic torque measurements have also been performed
on the optimally-doped material.

%
%
%
\begin{figure}[tb]
\includegraphics[width=2.7in]{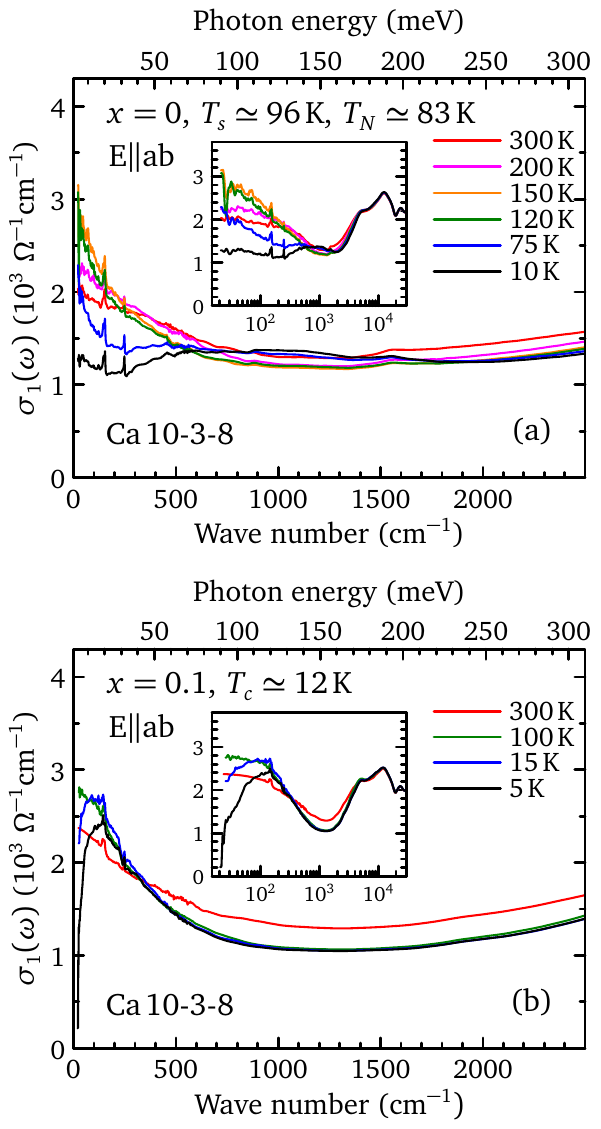}
\caption{The temperature dependence of the real part of the optical conductivity in the
\emph{a-b} planes of (CaFe$_{1-x}$Pt$_{x}$As)$_{10}\-$Pt$_{3}$As$_{8}$ for (a) $x=0$, and
(b) $x=0.1$; the insets show the optical conductivity at several temperatures over a wide
frequency range.}
\label{fig:sigma}
\end{figure}
%
%

%
%
\section{Results and Discussion}
%
%
%
The temperature dependence of the real part of the in-plane optical conductivity [$\sigma_{1}(\omega)$]
is shown in the infrared region for the undoped and doped compounds in Figs.~\ref{fig:sigma}(a) and
\ref{fig:sigma}(b), respectively; the conductivity is shown over a much broader frequency range in
the insets.  Several sharp features in the conductivity are observed at $\simeq 150$ and 250~cm$^{-1}$,
which are attributed to infrared-active lattice vibrations.   The extrapolated values for the dc resistivity
[$\sigma_{1}(\omega\rightarrow 0) \equiv \sigma_{dc}$, circles in Figs.~\ref{fig:resis}(b) and~\ref{fig:resis}(c)]
are essentially identical to the resistivity, indicating the excellent agreement between optics and transport
measurements.

The FeSCs are multiband materials; a minimal description consists of two electronic subsystems
using the so-called two-Drude model \cite{Wu2010} with the complex dielectric function
$\tilde\epsilon=\epsilon_1+i\epsilon_2$,
\begin{equation}
  \tilde\epsilon(\omega) = \epsilon_\infty - \sum_{j=1}^2 {{\omega_{p,D;j}^2}\over{\omega^2+i\omega/\tau_{D,j}}}
    + \sum_k {{\Omega_k^2}\over{\omega_k^2 - \omega^2 - i\omega\gamma_k}},
  \label{eq:eps}
\end{equation}
where $\epsilon_\infty$ is the real part at high frequency.  In the first sum
$\omega_{p,D;j}^2 = 4\pi n_je^2/m^\ast_j$ and $1/\tau_{D,j}$ are the square of the
plasma frequency and scattering rate for the delocalized (Drude) carriers in the $j$th
band, respectively, and $n_j$ and $m^\ast_j$ are the carrier concentration and effective mass.
In the second summation, $\omega_k$, $\gamma_k$ and $\Omega_k$ are the position, width, and
strength of the $k$th vibration or bound excitation.  The complex conductivity is
$\tilde\sigma(\omega) = \sigma_1 +i\sigma_2 = -2\pi i \omega [\tilde\epsilon(\omega) -
\epsilon_\infty ]/Z_0$ (in units of $\Omega^{-1}$cm$^{-1}$); $Z_0\simeq 377$~$\Omega$ is
the impedance of free space.  The model is fit to the real and imaginary parts of the
optical conductivity simultaneously using a non-linear least-squares technique.
%

%
%
%
\begin{figure*}[tb]
\includegraphics[width=5.75in]{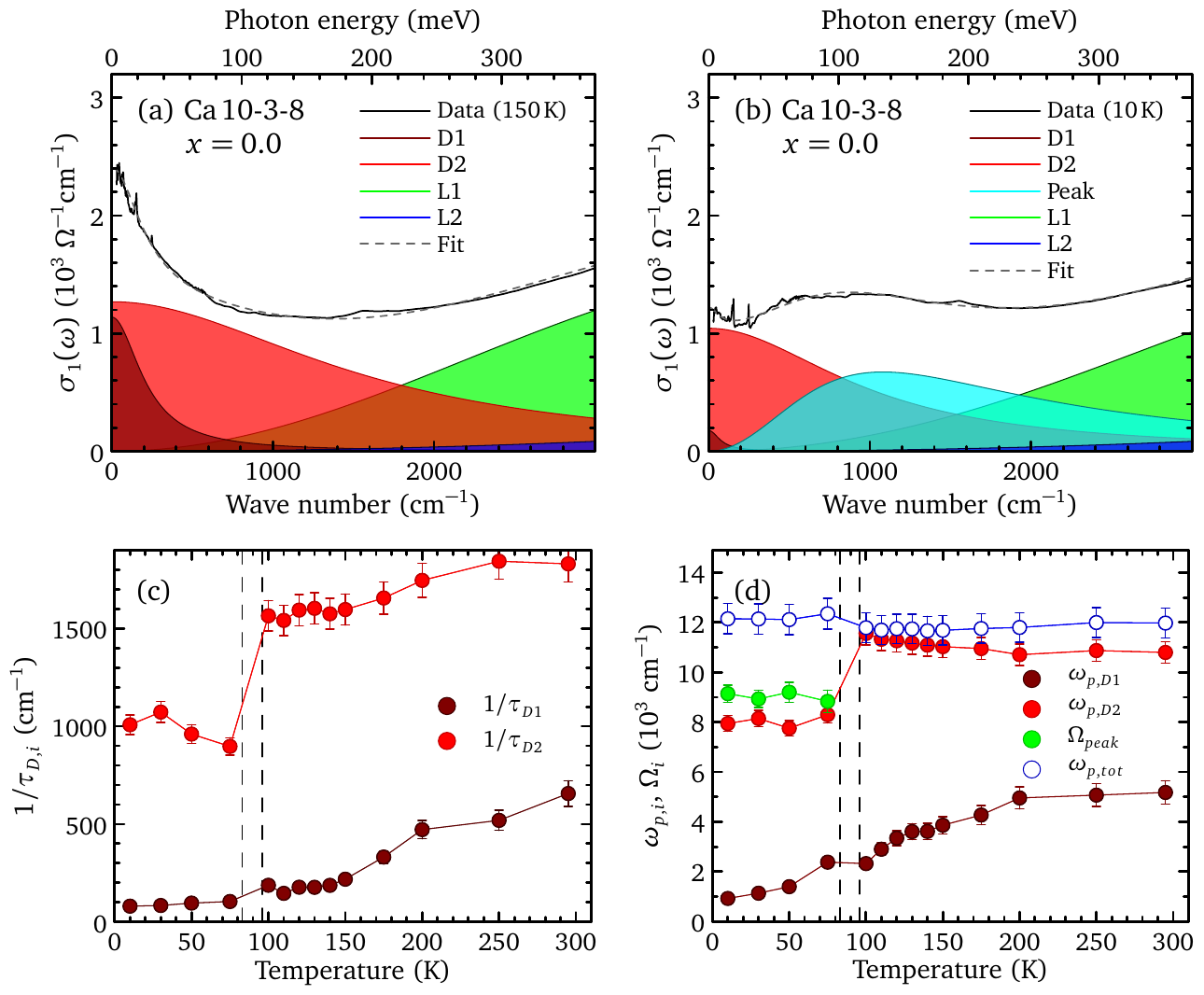}
\caption{The Drude-Lorentz model fits to $\sigma_{1}(\omega)$ of the Ca~10-3-8 undoped parent
compound at (a) 150~K and (b) 10~K, decomposed into the narrow (D1) and broad (D2) Drude components,
as well as several bound excitations; the peak that emerges below $T_s\simeq 96\,$K, $T_N\simeq 83\,$K
(dashed lines) has been fit using a Lorentzian line shape.
(c) The temperature dependence of the scattering rates for the narrow and broad Drude components.
(d) The plasma frequency for the narrow and broad Drude components, the oscillator strength of the
mid-infrared peak, and the total of all three contributions.}
\label{fig:parent}
\end{figure*}
%

%
%
\subsection{Parent compound ($\mathbf{x=0}$)}
At room temperature the optical conductivity in Fig.~\ref{fig:sigma}(a) has a metallic
character, with a Drude-like free carrier response superimposed on a flat, nearly incoherent
background.  The results of the two-Drude model fit at 150~K is shown in Fig.~\ref{fig:parent}(a);
the conductivity may be described by a free-carrier response consisting of a narrow Drude
term that reflects the coherent response, and a much stronger, broad Drude component that
corresponds to a nearly incoherent background; several Lorentzian oscillators are included
to describe the bound excitations (interband transitions) at higher energies
[Fig.~\ref{fig:sigma}(a)].
As the temperature is lowered, the low-frequency conductivity is suppressed and a broad
peak develops in the mid-infrared region.  The results of the model fit at 10~K are shown in
Fig.~\ref{fig:parent}(b); while the broad Drude has been reduced slightly in strength, the
narrow Drude is strongly suppressed and a strong peak centered at $\simeq 1000$~cm$^{-1}$ has
emerged.

The temperature dependencies of the scattering rates and the plasma frequencies of the narrow
and broad Drude terms are shown in Figs.~\ref{fig:parent}(c) and \ref{fig:parent}(d),
respectively.  As the temperature is reduced, the scattering rate for the broad Drude term
shows a weak temperature dependence; however, for $T\lesssim T_s, T_N$, it undergoes a dramatic
reduction from about $\simeq 1600$ to $\simeq 1000$~cm$^{-1}$.  In contrast, the scattering
rate for the narrow Drude term has a strong temperature dependence, decreasing from
$\simeq 660$~cm$^{-1}$ at room temperature to $\simeq 180$~cm$^{-1}$ at 100~K, below which
it decreases rapidly to $\simeq 80$~cm$^{-1}$ at low temperature.
The temperature dependence of the plasma frequencies tells a similar story.  As the temperature
is reduced, the plasma frequency of the broad Drude term is essentially constant; however, for
$T\lesssim T_s, T_N$ it undergoes a dramatic reduction, losing roughly 50\% of its strength
($\propto \omega_{p,D;i}^2$).  At the same time, a broad peak of roughly equal strength appears
in the mid-infrared region; as the temperature is further reduced, the strength of both features
remains unchanged.  The narrow Drude initially shows little temperature dependence, but below
about 200~K it begins to decrease uniformly with temperature, showing only a slight discontinuity
at the structural and magnetic transitions, ultimately losing over 90\% of its original strength
at low temperature.  Even though the coherent component is losing strength with decreasing
temperature, the commensurate decrease in the scattering rate results in a slight decrease
in the resistivity [Fig.~\ref{fig:resis}(b)], until $\simeq 150\,$K, below which the resistivity
begins to increase.

%
%
These trends may also be observed in the behavior of the spectral weight.  The spectral weight
is defined here as
\begin{equation}
  W(T) = \frac{Z_0}{\pi^2} \int_{\omega_a}^{\omega_b} \sigma_{1}(\omega,\,{T})\, d\omega ,
\end{equation}
over the $\omega_a - \omega_b$ interval.  As the temperature is lowered, spectral weight
is transferred from high to low frequency as the scattering rates decrease, shown in
Fig.~\ref{fig:weight}(a).  This trend is gradually reversed below $\simeq 150$~K with spectral
weight now transferred from below 600~cm$^{-1}$ to a broad peak centered at $\sim 1000$~cm$^{-1}$.

%
%
The $f$-sum rule requires that the sum of the squares of the plasma frequencies,
$\omega_{p,tot}^2=\omega_{p,D1}^2+\omega_{p,D2}^2+\Omega_{peak}^2$, should remain
constant.  This is indeed the case, as shown in Fig.~\ref{fig:parent}(d), so it
may be inferred that below $T_s$ and $T_N$, strength is transferred from both the
coherent and incoherent bands into the mid-infrared excitation.
This type of behavior is widely observed in most parent compounds of the FeSCs \cite{Nakajima2014,Homes2016}
and is attributed to the formation of a spin-density-wave-like (SDW) gap \cite{Hu2008} and
subsequent reconstruction of the Fermi surface~\cite{Yin11} resulting in low-energy interband
transitions that lie in the infrared region.  Interestingly, the coherent
component begins to lose strength well above the structural and magnetic transitions
[Figs.\ref{fig:parent}(d) and \ref{fig:weight}(a)].
The parent material is highly 2D \cite{Yuan2015}, the interlayer magnetic coupling is
very weak, and is easily destroyed by doping \cite{Xiang2012}.  Before three dimensional (3D)
long-range AFM order can be established, intralayer 2D short-range AFM fluctuations are
present \cite{Dai2012,Xu2016}.  As a result, the semiconducting-like behavior above $T_{N}$
may be regarded as the precursor to AFM order.

%
%
%

%
%
\subsection{Optimally-doped compound ($\mathbf{x=0.1}$)}
\begin{figure}[tb]
\includegraphics[width=2.5in]{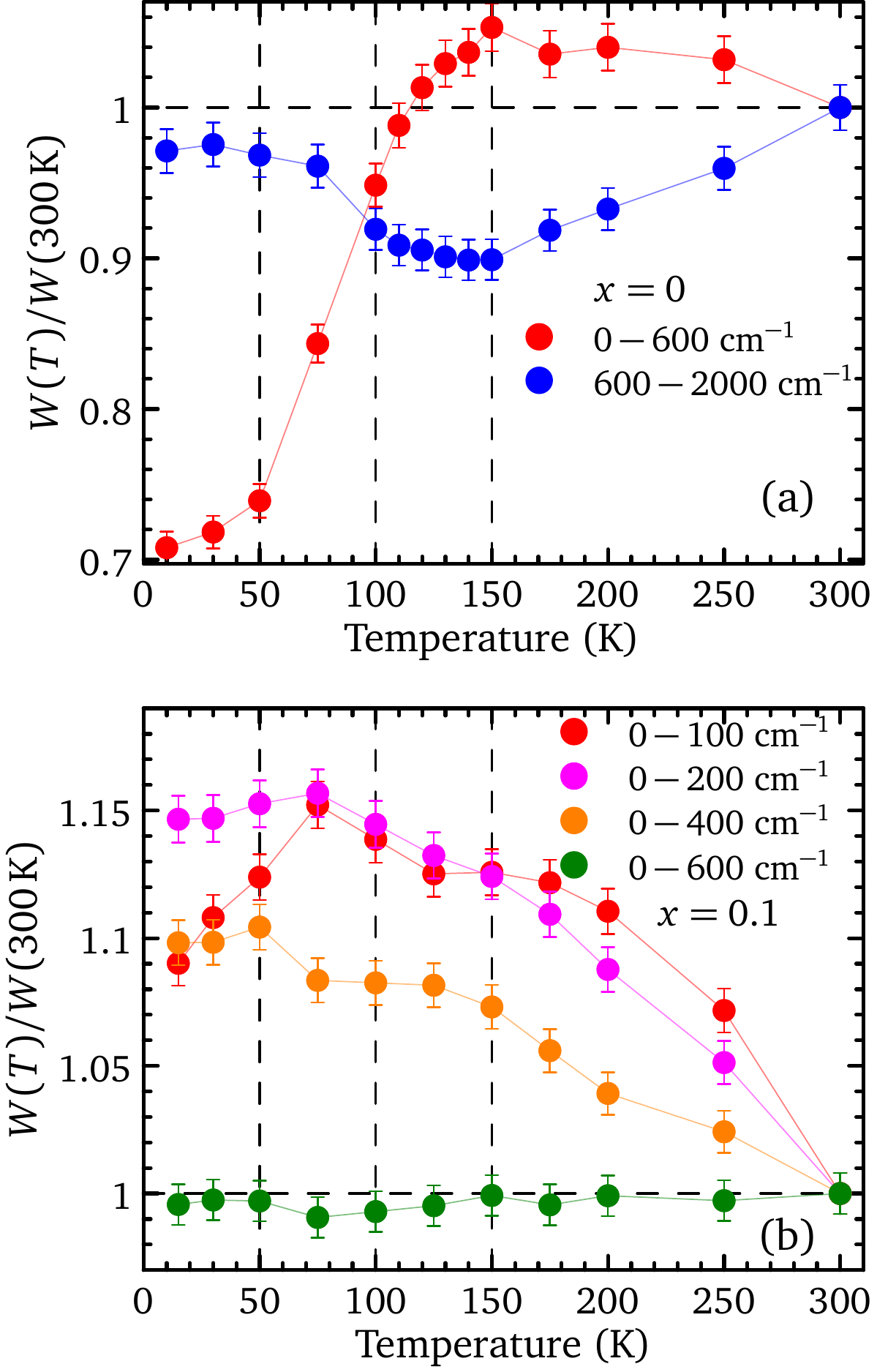}
\caption{The temperature dependence of the normalized spectral weight in the Ca 10-3-8
for the (a) undoped parent compound and (b) optimally-doped material over several different
frequency intervals. }
\label{fig:weight}
\end{figure}

In the optimally-doped sample, shown in Fig.~\ref{fig:sigma}(b), the optical conductivity exhibits
metallic behavior above $\sim 100$~K, although its temperature dependence is rather weak, with
spectral weight being gradually transferred from high to low frequency.
Below 100~K, the spectral weight below 100~cm$^{-1}$ is gradually suppressed while the
spectral weight below 600~cm$^{-1}$ remains constant, suggesting that the missing weight is
being transferred to the $100 - 600$~cm$^{-1}$ region [Fig.~\ref{fig:weight}(b)],
with the formation of a new absorption peak at $\simeq 120$~cm$^{-1}$.   At the same time, the
low-frequency conductivity is also decreasing, corresponding to the semiconducting-like response
below 100~K [Figs.~\ref{fig:resis}(c) and \ref{fig:sigma}(b)].
Upon entry into the superconducting state, the optical conductivity in the low-energy region
is almost completely suppressed, with $\sigma_{1}(\omega) \simeq 0$ below $\sim 20$~cm$^{-1}$,
signalling the opening of a nodeless superconducting energy gap~\cite{Yang2017}.

%
%
%
%

%
%
\subsubsection{Normal state}
\begin{figure*}[tb]
\includegraphics[width=5.75in]{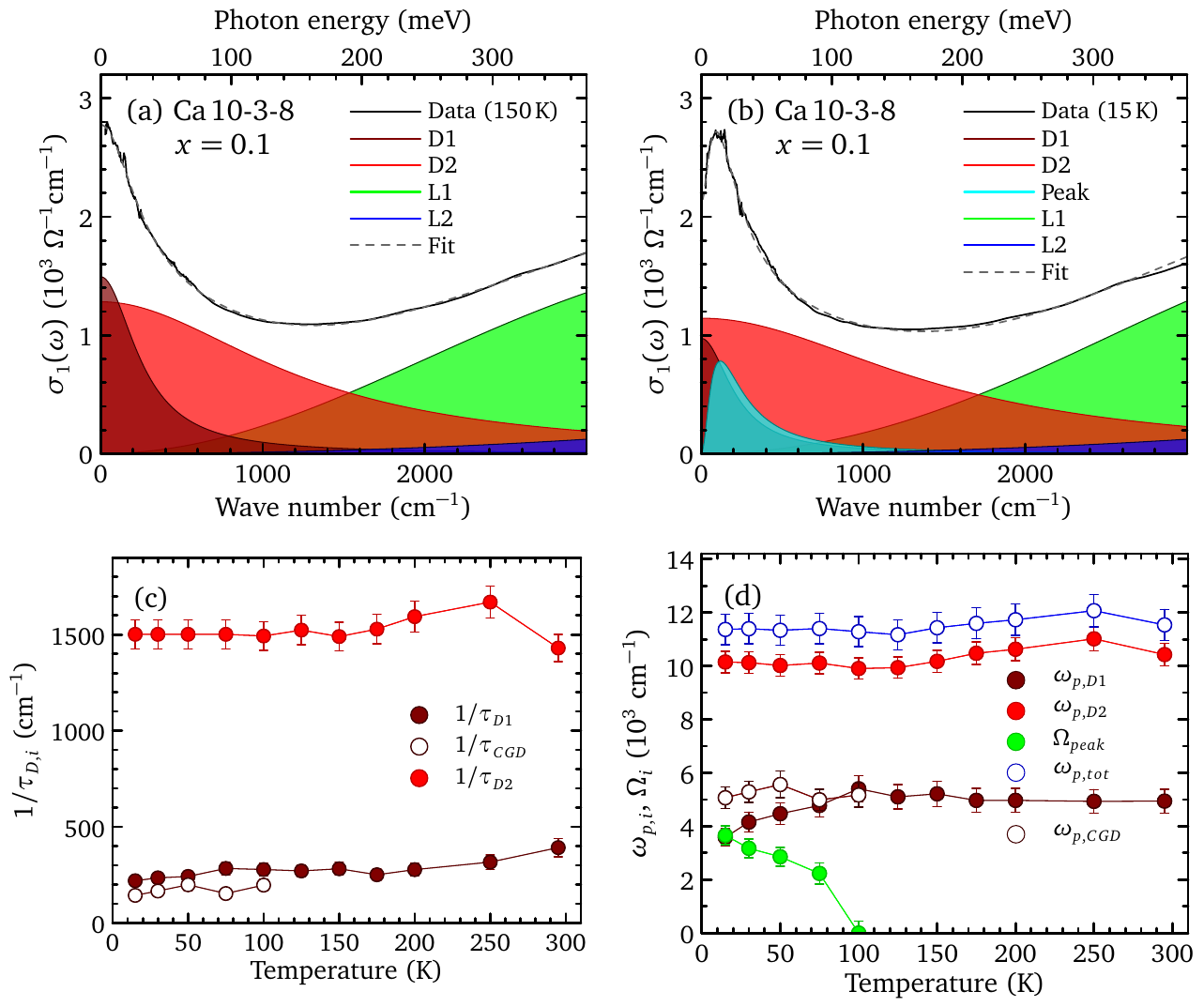}
\caption{The Drude-Lorentz model fits to $\sigma_{1}(\omega)$ of the optimally-doped Ca~10-3-8
at (a) 150~K and (b) 15~K, decomposed into the narrow (D1) and broad (D2) Drude components, as well
as several bound excitations; the peak that emerges at low temperature has been fit using a
Lorentzian line shape.
(c) The temperature dependence of the scattering rates for the narrow and broad Drude components
for the two-Drude model, as well as that of the classical generalized Drude model (CGD) below $T^\ast
\simeq 100\,$~K.
(d) The plasma frequency for the narrow and broad Drude components, the oscillator strength
of the emergent peak, and the total of all three contributions; in the CGD model $\omega_{p,CGD}
(T\leq T^\ast)\simeq \omega_{p,D1}(T > T^\ast)$.}
\label{fig:doped}
\end{figure*}

The results of the fits using the two-Drude model to the conductivity of the
optimally-doped sample at 150 and 15~K are summarized in Figs.~\ref{fig:doped}(a)
and \ref{fig:doped}(b), respectively.
As observed in the parent compound, at room temperature the conductivity may be
described by narrow and broad Drude terms; several Lorentzian oscillators
are included to describe the bound excitations (interband transitions) at higher energies
[Fig.~\ref{fig:sigma}(b)].  Below 100~K, the decrease in intensity of the narrow Drude
component is accompanied by the formation of a peak at $\simeq 120$~cm$^{-1}$
[Fig.~\ref{fig:doped}(b)], which has been fit using a Lorentzian line shape.  The
temperature dependencies of the scattering rates and plasma frequencies are shown in
Figs.~\ref{fig:doped}(c) and \ref{fig:doped}(d), respectively.
By tracking the strengths of the plasma frequencies of the Drude terms, we note that
below $T^\ast\simeq 100$~K the plasma frequency of the narrow Drude is suppressed
while the strength of the new peak is gradually enhanced [Fig.~\ref{fig:doped}(d)];
the broad Drude term displays little temperature dependence above or below $T^\ast$.
The conservation of spectral weight again requires that $\omega_{p,tot}^2 =
\omega_{p,D1}^2+\omega_{p,D2}^2+\Omega_{peak}^2$ should remain constant, which
is indeed the case in Fig.~\ref{fig:doped}(d).  Thus, below $T^\ast$, some of
the coherent response (narrow Drude) is transferred to the new peak in the optical
conductivity, resulting in a reduced $\sigma_{dc}$ and semiconducting-like behavior.

%
%
%

%
%
The evolution of the low-energy peak may also be described using a simple
classical generalization of the Drude (CGD) formula in which the faction of
the carriers velocity that is retained after a collision \cite{Smith2001}.
While many collisions may be considered, in the single-scattering approximation
the complex conductivity is written as
\begin{equation}
  \tilde\sigma(\omega)= \left(\frac{2\pi}{Z_0}\right) \frac{\omega_p^2\tau}{(1-i\omega\tau)}
  \left[ 1+\frac{c}{(1-i\omega\tau)} \right],
  \label{eq:smith}
\end{equation}
where $c$ is the persistence of velocity that is retained for a single collision.
This model has the interesting attribute that for $c=0$ a simple Drude is
recovered, while for $c=-1$ the carriers are completely localized in the form of
a Lorentzian oscillator with a peak at $\omega\tau=1$, width $2/\tau$, and an
oscillator strength that is identical to the plasma frequency (Supplemental Material).
The real and imaginary parts of the optical conductivity have been fit between 15 and
125~K using the two-Drude model with the provision that the narrow Drude component
is replaced by the expression in Eq.~(\ref{eq:smith}).  At 125~K, the fit is identical
to that of the two-Drude model, returning $c=0$ (pure Drude).  Fits below 100~K reveal
an increasingly negative value for $c=-0.26$, -0.49, -0.41, -0.51, and $-0.61\pm 0.05$
at 100, 75, 50, 30, and 15~K, respectively.  Interestingly, the plasma frequency
for the narrow Drude is now roughly constant, with $\omega_{p,D1}\simeq
5200\pm 500$~cm$^{-1}$, which is essentially identical to the values returned
from the two-Drude model at and above 125~K [Fig.~\ref{fig:doped}(d)]; this
indicates that the response of both the localized and free carriers is now
incorporated into a single plasma frequency.
The scattering rate is also slightly lower at low temperatures than the values
obtained using the two-Drude model [Fig.~\ref{fig:doped}(c)].  The values for
the broad Drude term are unchanged.   While the overall quality of the
fits is indistinguishable from the two-Drude model, it is remarkable that the
introduction of a single new parameter to the narrow Drude band allows both the
position and strength of the low-energy peak to be described quite well, indicating
that this peak likely arises from carrier localization due to scattering.
In addition, the value of $c\simeq -0.6$ at 15~K is consistent with the results
of the two-Drude with a Lorentzian that indicate that the spectral weight of the
narrow Drude component is split more or less equally between free and localized
carriers at low temperature.
%
%
These results are in good agreement with recent angle-resolved photoemission
spectroscopy (ARPES) measurements, which show a clear decrease in the size of the
hole pockets below $T^\ast$ (Supplementary Fig.~S2).

%
%
\subsubsection{Superconducting state}
\begin{figure}[tb]
\includegraphics[width=2.5in]{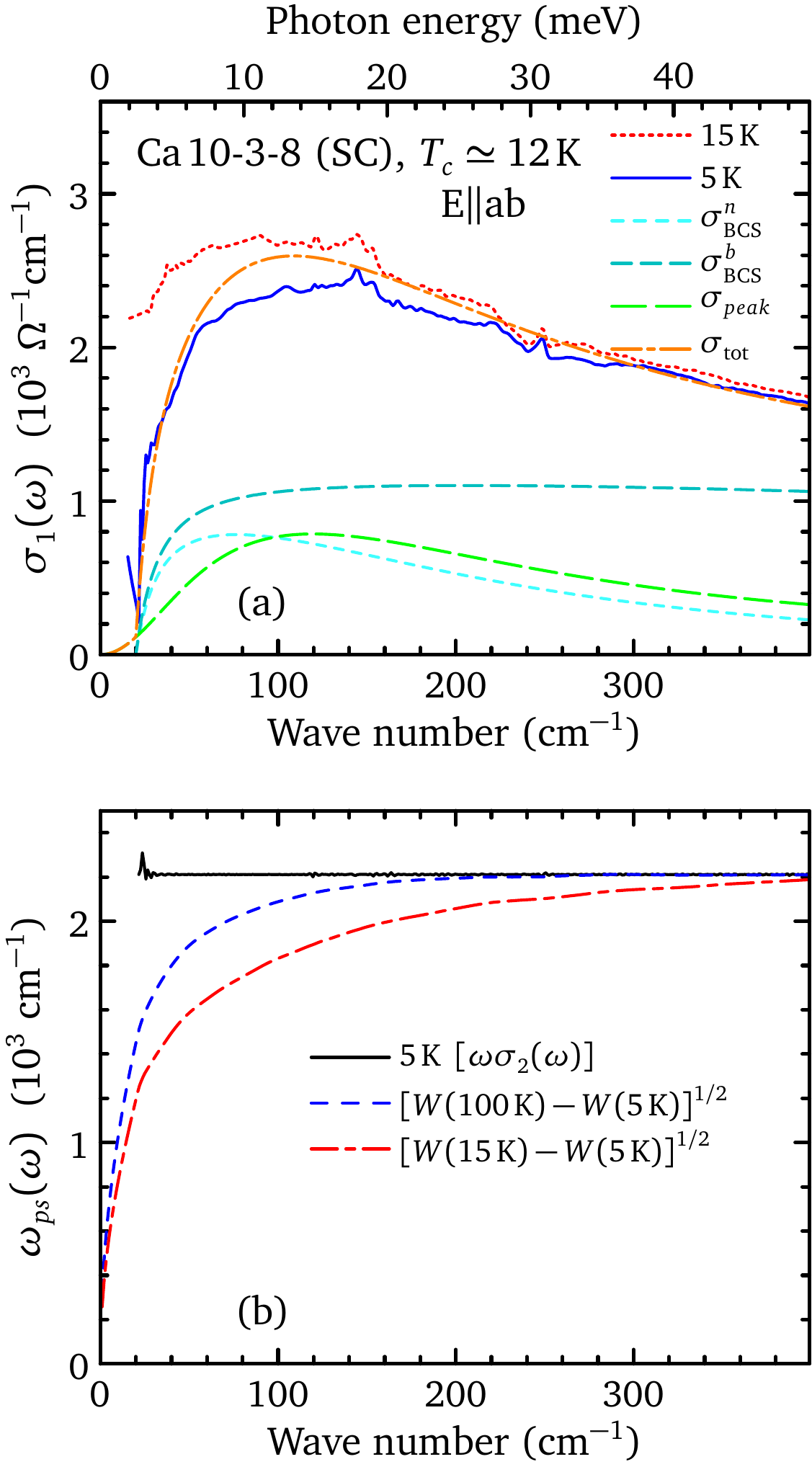}
\caption{(a) The real part of the optical conductivity of optimally-doped
Ca 10-3-8 just above (dotted line) and below (solid line) $T_{c}$. The
short dashed lines are the gapped spectrum of excitations for the narrow
and broad Drude bands with $2\Delta_{1,2} \simeq 20$~cm$^{-1}$; the long dashed line is the
contribution of the emergent peak.  The linear combination of these contributions
(dashed-dot line) reproduces the data quite well, except in the region of
peak.
(b) The superfluid weight (solid black line) obtained from the imaginary part of
$\omega\sigma_{2}(\omega)$ (see Ref.~\onlinecite{Homes2004} for details). The red
and blue dashed lines are obtained from the FGT sum rule [Eq.~(\ref{eq:FGT})].}
\label{fig:super}
\end{figure}

Although there is semiconducting-like response in the normal state, below $T_{c}
\simeq 12$~K, the resistivity drops to zero; such a semiconducting-like to
superconducting transition is unusual in iron-based superconductors.  A clear signature
of superconducting transition is observed in the reflectivity [Supplementary Fig.~S1(b)].
In the real part of the optical conductivity, the spectral weight below $\simeq 20$~cm$^{-1}$
is totally suppressed, indicating the opening of a nodeless superconducting energy gap.
The Mattis-Bardeen formalism is used to describe the gapping of the spectrum of excitations in
the superconducting state \cite{Zimmerman91,Dressel-Book}.  The real part of the optical
conductivity is shown just above $T_c$ by the dotted line in Fig.~\ref{fig:super}(a); this curve
is described by narrow ($\omega_{p,D1}\simeq 3580$~cm$^{-1}$, $1/\tau_{D1}\simeq 220$~cm$^{-1}$)
and broad ($\omega_{p,D2}\simeq 10\,150$~cm$^{-1}$, $1/\tau_{D2}\simeq 1500$~cm$^{-1}$) Drude
components, as well as a low-energy bound excitation ($\omega_0\simeq 120$~cm$^{-1}$,
$\gamma_0\simeq 285$~cm$^{-1}$, and $\Omega_0\simeq 3660$~cm$^{-1}$).  The data below
$T_c$ at $\simeq 5\,$K is shown by the solid line; despite the presence of multiple
bands and a weak shoulder at $\simeq 30$~cm$^{-1}$, we have chosen to simplify the analysis
and model the data with a single superconducting energy scale for both bands,
$2\Delta_1 = 2\Delta_2 \simeq 20\pm 4$~cm$^{-1}$ ($\simeq 2.5\pm 0.4$~meV).
Note that $1/\tau_{D1} > 2\Delta_{1,2}$, and $1/\tau_{D2}\gg 2\Delta_{1,2}$, placing this
material in the dirty limit \cite{Homes2015}.  The gapped spectrum of excitations for the
two Drude bands, as well as the contribution from the emergent peak, are shown by the
dashed lines in Fig.~\ref{fig:super}(a).
The linear combination of all three contributions reproduces the data quite well,
except in the region of the peak.  It should be noted that the normal-state
values are not refined to fit the data below $T_c$.  The 5~K data may be more accurately
reproduced by decreasing the intensity of the peak by about 20\%, suggesting that some
of the spectral weight of this feature has collapsed into the condensate.
The gap ratio ${2\Delta_{1,2}}/{k_{\rm B}T_{c}} \simeq 2.4$ falls below the BCS value
of 3.5, placing this material in the BCS weak-coupling limit
%
%

%
%

The formation of superconducting energy gap(s) below $T_c$ results in the loss of low-frequency
spectral weight that collapses into the superfluid condensate; the strength of the condensate may
be estimated in one of two ways.  The complex conductivity for the superfluid response may be
expressed as~\cite{Dordevic2002,Hwang2007}
\begin{equation}
  \tilde\sigma_{s}(\omega)= \sigma_{s1}+\emph{i}\,\sigma_{s2}(\omega) =
    \frac{\pi^2}{Z_{0}}\omega_{ps}^{2}\delta(0)+\frac{\emph{i} 2\pi\omega_{ps}^{2}}{Z_{0}\omega},
  \label{eq:sigma}
\end{equation}
where $\omega_{ps}^{2}=4\pi n_{s}e^{2}/m^\ast$ represents the superconducting plasma frequency,
$n_{s}$ is the superconducting carrier density, and $m^\ast$ is an effective mass.  Thus, from
the imaginary part $2\pi \omega_{ps}^2 \simeq Z_0\omega\sigma_{s2}(\omega)$.  Alternatively, the
difference between the low-frequency optical conductivity just above and below $T_c$, the
so-called ``missing area'', can be analyzed using the Ferrel-Glover-Tinkham (FGT) sum
rule \cite{Ferrell1958,Tinkham1959}:
\begin{equation}
  \frac{Z_{0}}{\pi^{2}}\int_{0^{+}}^{\omega}[\sigma_{1}(\omega^\prime,T\gtrsim T_{c}) -
    \sigma_{1}(\omega^\prime,T\ll T_{c})] d\omega^\prime = \omega_{ps}^{2},
  \label{eq:FGT}
\end{equation}
where the cutoff frequency $\omega$ is chosen so that the integral converges smoothly.
Both methods yield similar values of $\omega_{ps}\simeq 2\,110\pm 200$~cm$^{-1}$, shown in
Fig.~\ref{fig:super}(b), resulting in a penetration depth of $\lambda_{0}= 7\,500\pm 600$~{\AA},
in agreement with previous $\mu$SR measurements~\cite{Surmach2015}.
%
%

%
%
While about half of the spectral weight of the narrow Drude component has been transferred to the
far-infrared absorption peak [Fig.~\ref{fig:doped}(d)], as Fig.~\ref{fig:super}(a) indicates, this
peak is also suppressed below $T_{c}$.  A key question is: What becomes of these localized electrons?
To address this question, we have applied the FGT sum rule by taking the difference in the optical
conductivity between 15 and 5~K, and 100 and 5~K.
From the results shown in Fig.~\ref{fig:super}(b), we notice that the superfluid stiffness
calculated with respect to 15 and 5~K converge at $\omega \simeq 400$~cm$^{-1}$ ($\sim 50$~meV).
However, between 100 and 5~K the integral converges much more quickly ($\omega\lesssim 200$~cm$^{-1}$),
a very unusual situation.  If only the Drude components condense into the superfluid, the results
calculated between 15 and 5~K would converge more quickly, because the Drude component is narrower
at low temperature [Figs.~\ref{fig:doped}(c) and \ref{fig:doped}(d)].  Such anomalous behavior
suggests that there is an extra component in the $100- 400$~cm$^{-1}$ region that contributes to
the superfluid below $T_c$.  This implies that the newly-formed peak below 100~K contributes
to the superfluid condensate.  Understanding the origin of this peak may provide insight
into the unconventional pairing in this material.

In order to further understand the relation between the semiconducting-like
behavior and superconductivity, we performed a magnetic torque measurement
on the optimally-doped sample (details are provided in the Supplementary Material).
In Fig.~\ref{fig:torque}(a), we observe that, below $T^{*}$, the torque $\tau_{0}$ starts to
deviate from the high-temperature \emph{T}-linear behavior and $|\chi_{c}-\chi_{ab}|$ increases
with decreasing $H$ [Fig.~\ref{fig:torque}(b)].  Both types of behavior indicate a non-linear
susceptibility \cite{Xiao2014}.  Approaching $T_{c}$, this non-linearity appears to diverge
in the zero-field limit, suggesting that this behavior may be related to superconducting
fluctuations \cite{Kasahara2016}.
Inelastic neutron scattering and nuclear magnetic resonance have both observed
evidence for preformed Cooper pairs in Ca 10-3-8~\cite{Surmach2015}, and the minimum
in the temperature-dependent Seebeck coefficient of Pt-doped material can be understood in
terms of either preformed pairs or the phonon drag effect \cite{Ni2011}.
A difficulty with the notion of preformed pairs is that they are typically observed in
strong-coupling systems, in which the coherence length is comparable with the inter-particle
distance~\cite{Kasahara2016}; however, optimally-doped Ca 10-3-8 is in the BCS weak-coupling limit.
Furthermore, in cuprates and FeSe, superconducting fluctuations always result in a decrease in
the resistance \cite{Popcevic2018}; the semiconducting-like behavior is anomalous.

%
%
\begin{figure}[tb]
\includegraphics[width=2.55in]{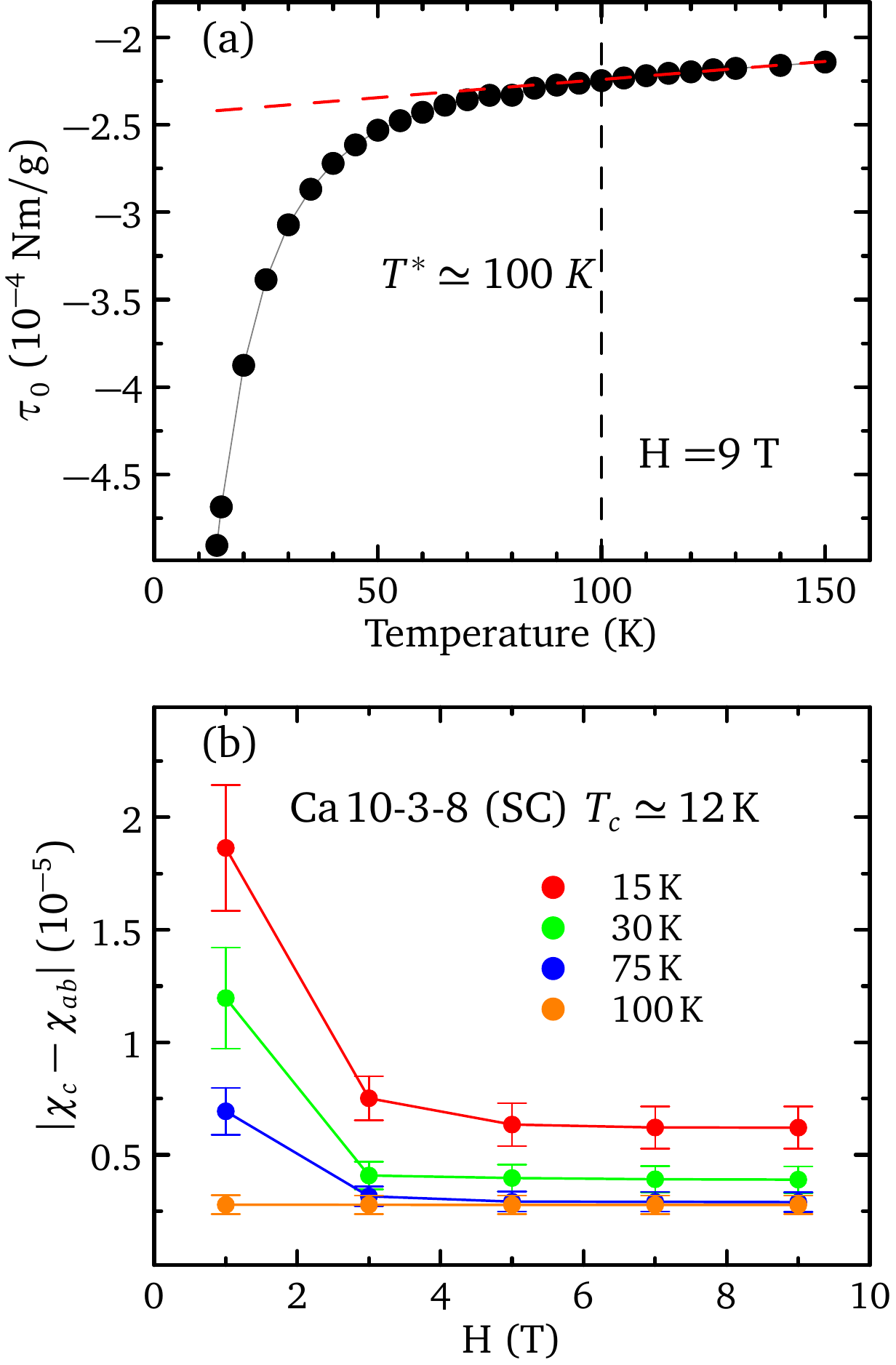}
\caption{(a) The temperature-dependent out-of-plane torque $\tau_{0} =
\frac{1}{2}\mu_{0}(\chi_{c}-\chi_{ab})H^{2}$ for the optimally-doped Ca~10-3-8 with fixed
magnetic field (9~T); $\chi_{c}$ and $\chi_{ab}$ are magnetic susceptibilities along the
\emph{c} and \emph{a} axis, respectively. The dashed line is linear fit to the high-temperature data.
(b) The field-dependent $|\chi_{c}-\chi_{ab}|$ at different temperatures.}
\label{fig:torque}
\end{figure}

%
%
\subsection{Localization and magnetism}
%
%
The substitution of Pt for Fe into the Fe--As planes of (CaFe$_{1-x}$Pt$_{x}$As)$_{10}$Pt$_{3}$As$_{8}$
induces superconductivity; however, it also results in the introduction of disorder sites that
can lead to strong scattering and the localization of free-carriers.  In systems that display
activated behavior, the interplay between localization and superconductivity is of considerable
interest \cite{Ma1985,Valles1989}.  In YBa$_2$(Cu$_{1-x}$Zn$_x$)$_4$O$_8$, the substitution
of Zn for Cu led to the dramatic reduction of $T_c$ and the appearance of a peak in the
optical conductivity at $\simeq 120$~cm$^{-1}$ that was attributed to quasiparticle
localization \cite{Basov1998}.  In the cuprates, non-magnetic Zn is thought to act as
a magnetic impurity.  A low-energy peak in in the optical conductivity was also observed
when the magnetic impurities Mn and Cr were substituted for Fe in BaFe$_2$As$_2$ \cite{Kobayashi2016}.
However, in (CaFe$_{1-x}$Pt$_{x}$As)$_{10}$Pt$_{3}$As$_{8}$, the Pt$^{4+}$ atoms in the
Fe--As layers are nonmagnetic and a low-energy peak has never been observed in iron-based
superconductors with nonmagnetic impurities~\cite{Nakajima2010}.
We note that the semiconducting-like behavior is not distinct to Pt-doped Ca 10-3-8;
under pressure, the stoichiometric parent compound also shows semiconducting-like
behavior, right up to the point at which it becomes a superconductor.  The similarity
between the phase diagrams of Pt-doped Ca10-3-8, and Ca 10-3-8 under pressure,
indicates the intrinsic nature of the semiconducting-like behavior.  Moreover,
the low-energy peak observed in Pt-doped Ca 10-3-8, has also been seen in La-doped
Ca$_{8.5}$La$_{1.5}$(Pt$_3$As$_8$)(Fe$_2$As$_2$)$_5$ (out-of-plan doping) \cite{Seo2017},
and in stoichiometric (CaFeAs)$_{10}$Pt$_4$As$_8$ \cite{Seo2018}.
The 2D nature of these materials  will greatly increase the importance of spin fluctuations,
suggesting that they may be playing a prominent role in the in-plane transport properties.
It is therefore likely that the localization peak observed in this work arises from strong
scattering due to AFM fluctuations rather than impurity scattering.

%
%

Although AFM order competes with superconductivity, spin fluctuations have been proposed
as a possible pairing mechanism in the high-temperature superconductors \cite{Moriya2000}.
The torque magnetometry results [Fig.~\ref{fig:torque}(a)] indicate that the onset for
superconducting fluctuations occur well above $T_c$.  This type of behavior has been
observed in many high-temperature superconductors \cite{Li2010,Cyr2018}, and it has
been suggested that these fluctuations may be attributed to the inhomogeneous nature
(either structural or electronic) of these materials \cite{Pelc2018}.  Thus, it may be the
case that the putative normal-state is an effective medium in which the superconducting regions
are embedded in a poorly-conducting matrix with either strong spin fluctuations or magnetic order
(i.e., incommensurate SDW); below $T_c$, phase coherence is established across the
different superconducting regions and a bulk superconducting transition is observed.  The
global onset of superconductivity would naturally suppress the magnetic fluctuations (order)
and the scattering attributed to it, leading to a reduction in the size of the localization peak.
The carriers that are no longer localized due to strong scattering would then be allowed to collapse
into the condensate, a result that is consistent with the observed transfer of spectral weight
from the peak into the condensate below $T_c$.

\section{Summary}
To conclude, the temperature dependence of the in-plane optical properties of
(CaFe$_{1-x}$Pt$_{x}$As)$_{10}$Pt$_{3}$As$_{8}$ has been examined for the
undoped ($x=0$) parent compound with $T_s\simeq 96$~K and $T_N\simeq 83$~K, and
the optimally-doped ($x=0.1$) superconducting material, $T_c\simeq 12$~K.
At room temperature, the optical conductivity of both materials may be described
by the two-Drude model.  In the parent compound, below $T_s$ and $T_N$ the broad Drude
component narrows and decreases dramatically in strength, behavior which is also
observed in the narrow Drude component.  The missing spectral weight is transferred
to a broad peak at $\simeq 1000$~cm$^{-1}$, which is attributed to a low-energy
interband transition that originates from the Fermi surface reconstruction driven
by the structural and magnetic transitions.  The semiconducting-like behavior
originates from short-range magnetic fluctuations that could be regarded as the
precursor to AFM order.
In the optimally-doped material, the broad Drude term shows little temperature
dependence, but the scattering rate in the narrow Drude component has a weak
temperature dependence.  Below $T^\ast \simeq 100$~K, the narrow Drude loses
strength at the same time a localization peak at $\simeq 120$~cm$^{-1}$ emerges.
A classical generalization of the Drude model reproduces the
position and strength of the low-energy peak, indicating that it originates via
a localization process.
Torque magnetometry detects a diamagnetic signal well above $T_c$, which is attributed
to SC fluctuations.  Below $T_c$ magnetic fluctuations (order) are suppressed,
resulting in a decrease in localization, allowing spectral weight from this peak to
be transferred into the superconducting condensate.  These results indicate an intimate
relationship between magnetism and superconductivity in this material.

%
%
%

\begin{acknowledgments}
We thank Liling Sun, Jimin Zhao, Weiguo Yin, Yilin Wang, Hu Miao, Peter Johnson for useful discussions.
Work at Chinese Academy of Science was supported by NSFC (Project No. 11374345 and No. 91421304)
and  MOST (Project No. 2015CB921303 and No. 2015CB921102).
J.H. acknowledges the financial support from the National Research Foundation of Korea
(NRFK Grant No. 2017R1A2B4007387).
Work at Brookhaven National Laboratory was supported by the Office of Science, U.S. Department
of Energy under Contract No. DE-SC0012704.
\end{acknowledgments}
%
%
%
%

%

%
%

\newpage \ \\

\newpage
\vspace*{-2.0cm}
\hspace*{-2.5cm} {
  \centering
  \includegraphics[width=1.2\textwidth]{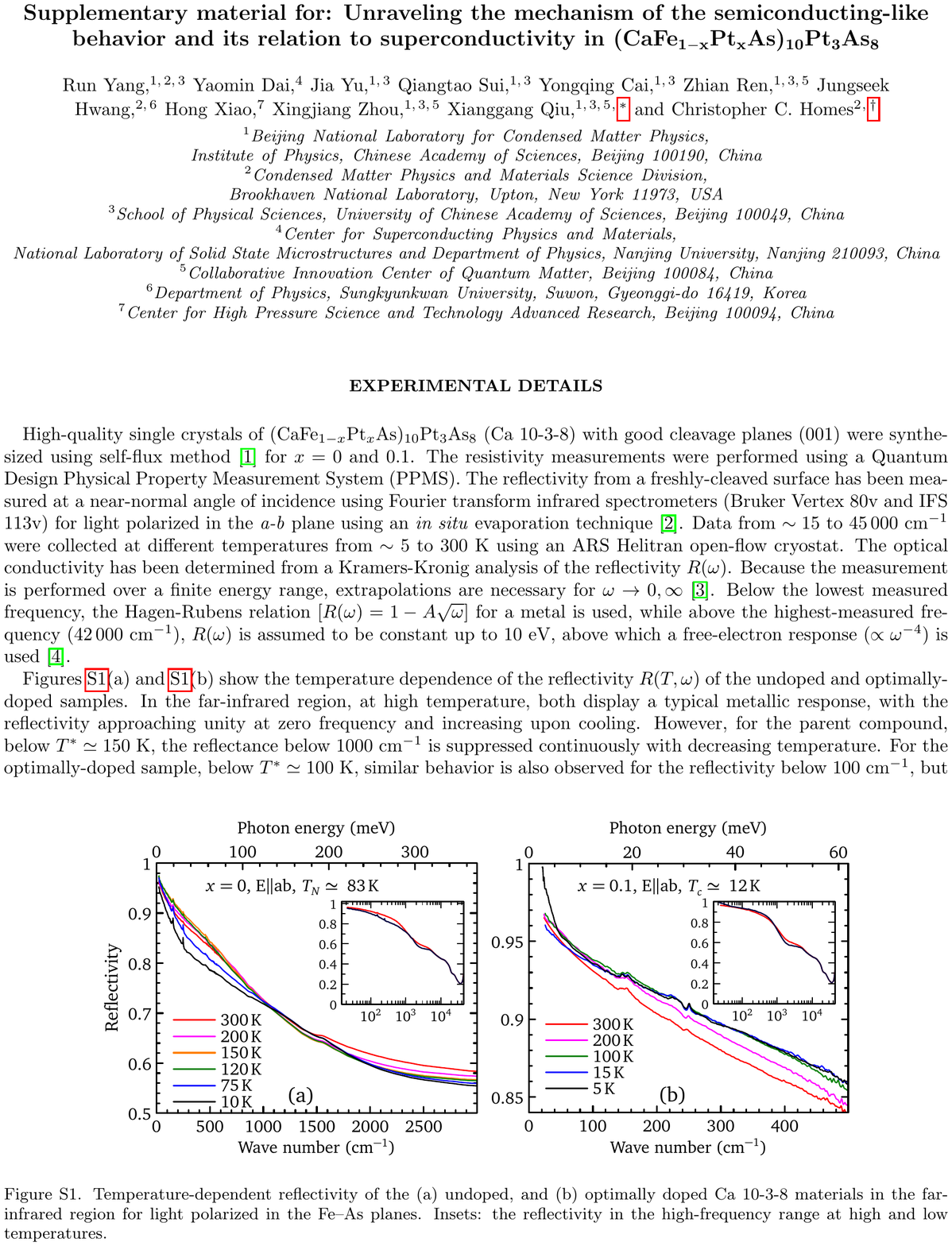} \\
  \ \\
}

\newpage
\vspace*{-2.0cm}
\hspace*{-2.5cm} {
  \centering
  \includegraphics[width=1.2\textwidth]{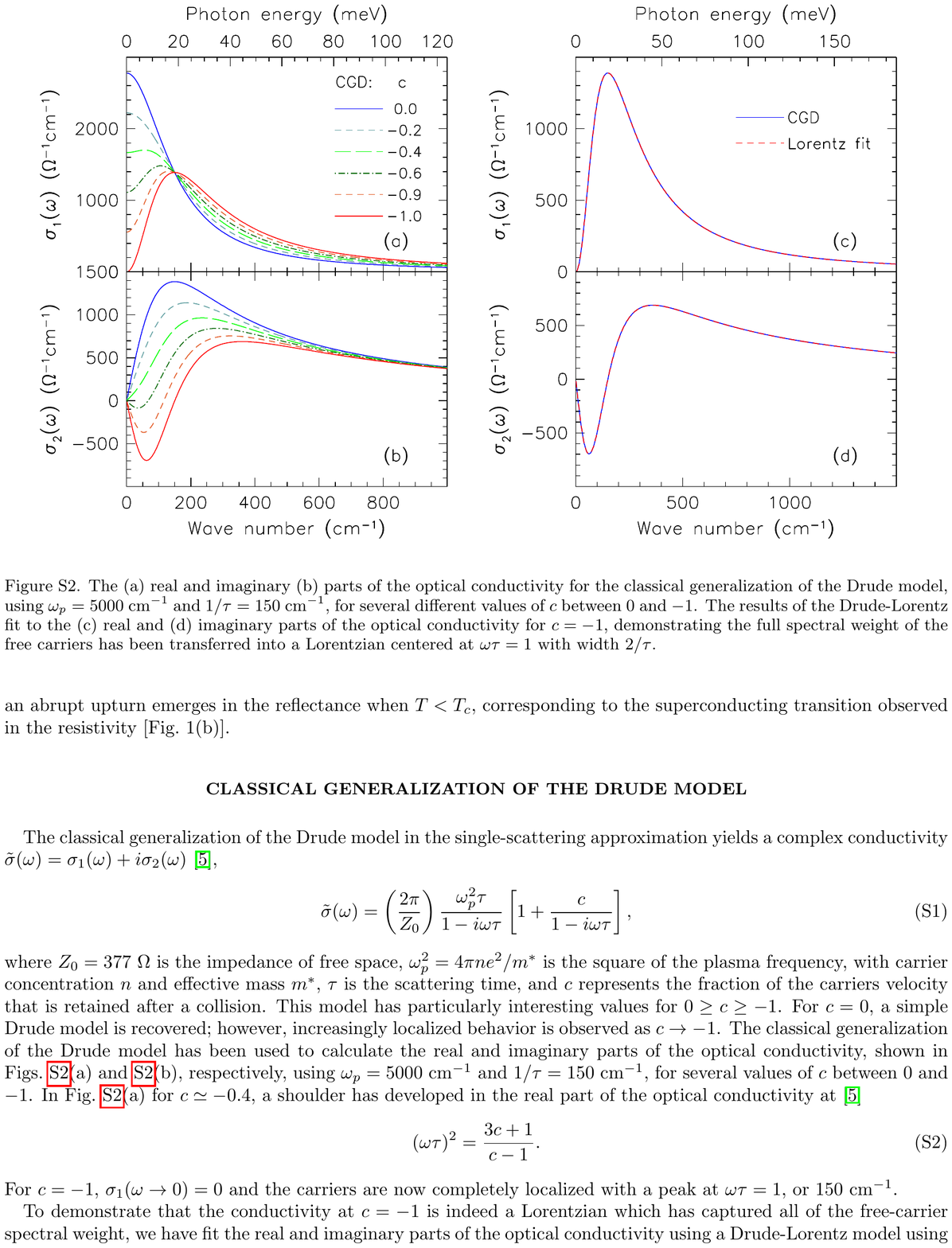} \\
  \ \\
}

\newpage
\vspace*{-2.0cm}
\hspace*{-2.5cm} {
  \centering
  \includegraphics[width=1.2\textwidth]{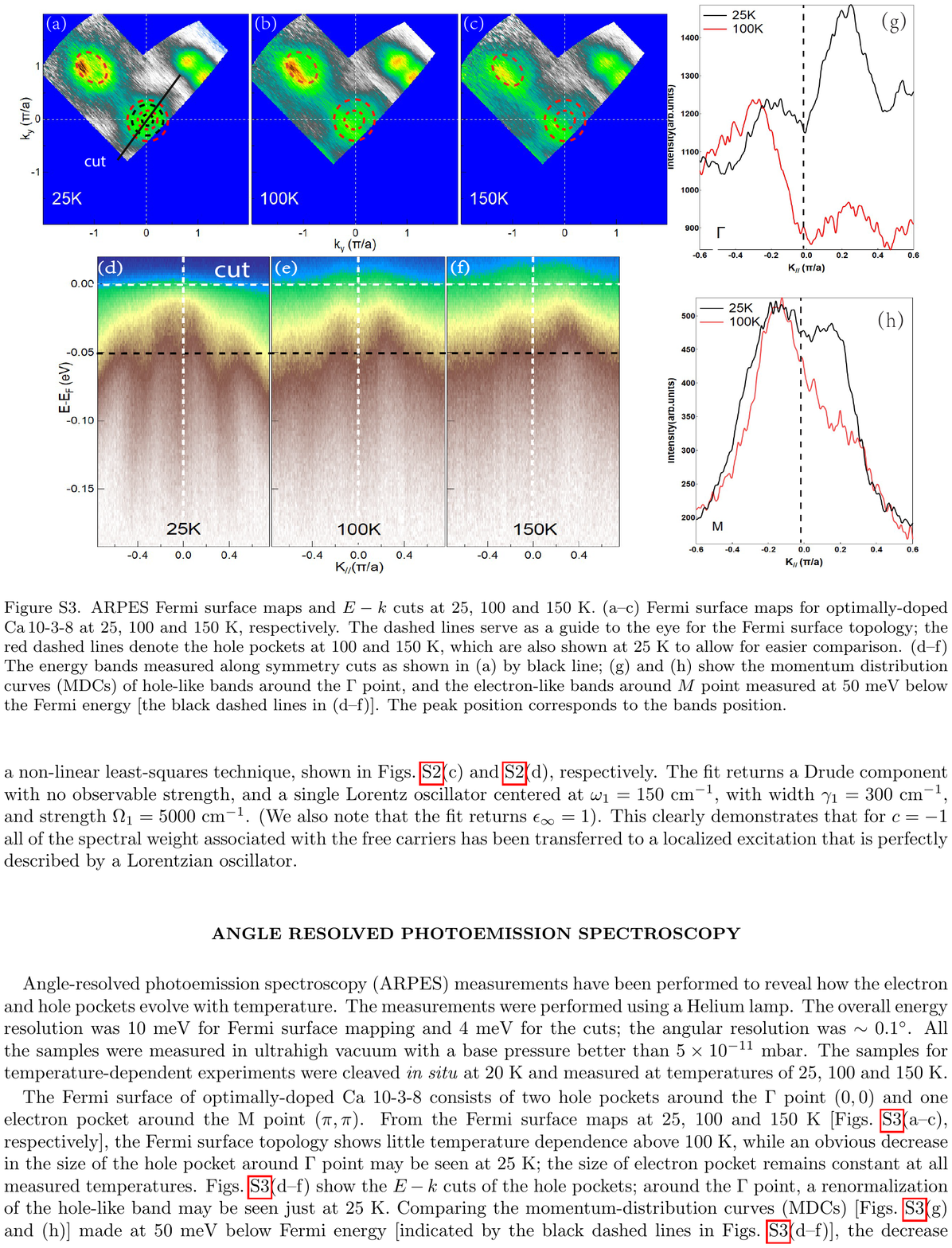} \\
  \ \\
}

\newpage
\vspace*{-2.0cm}
\hspace*{-2.5cm} {
  \centering
  \includegraphics[width=1.2\textwidth]{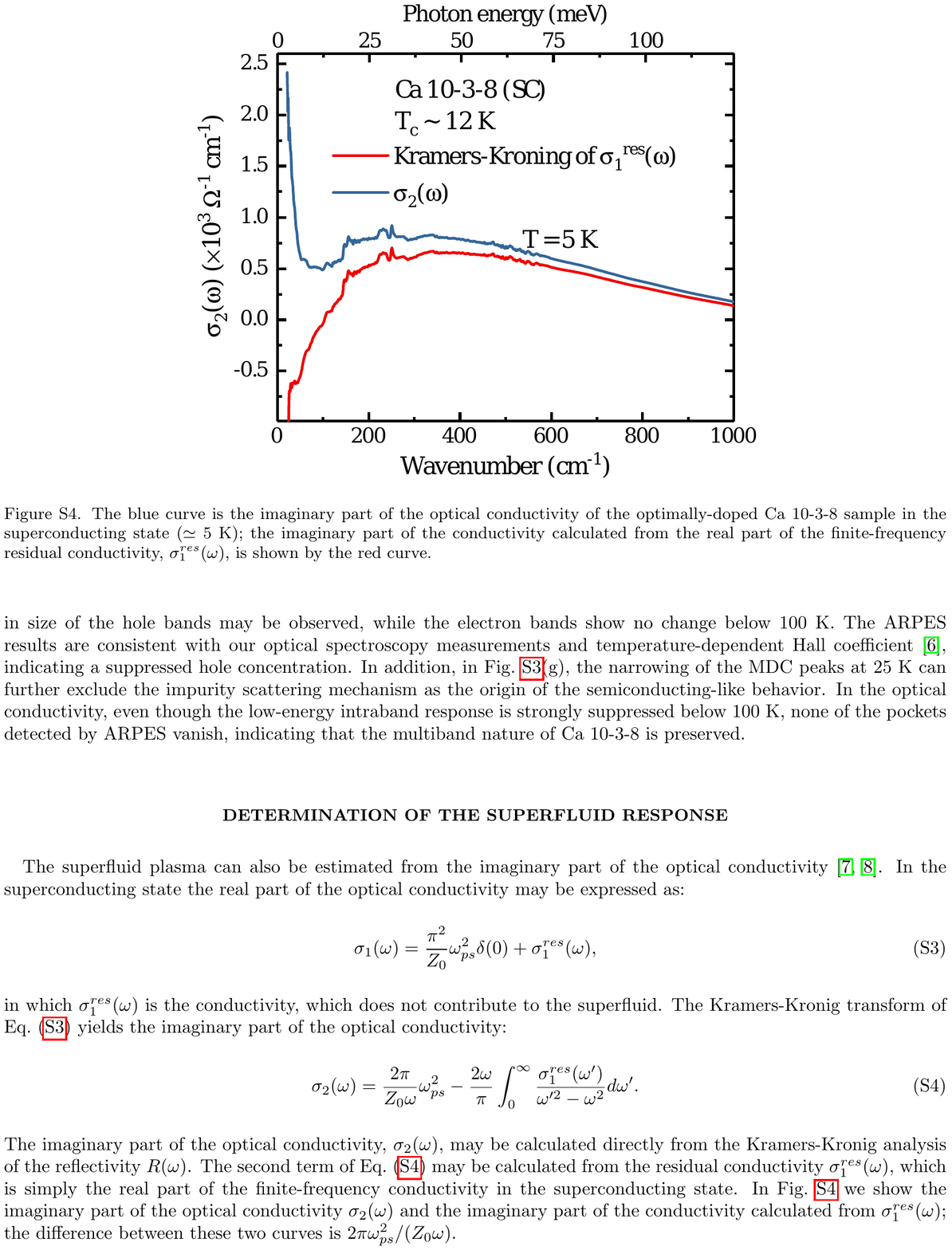} \\
  \ \\
}

\newpage
\vspace*{-2.0cm}
\hspace*{-2.5cm} {
  \centering
  \includegraphics[width=1.2\textwidth]{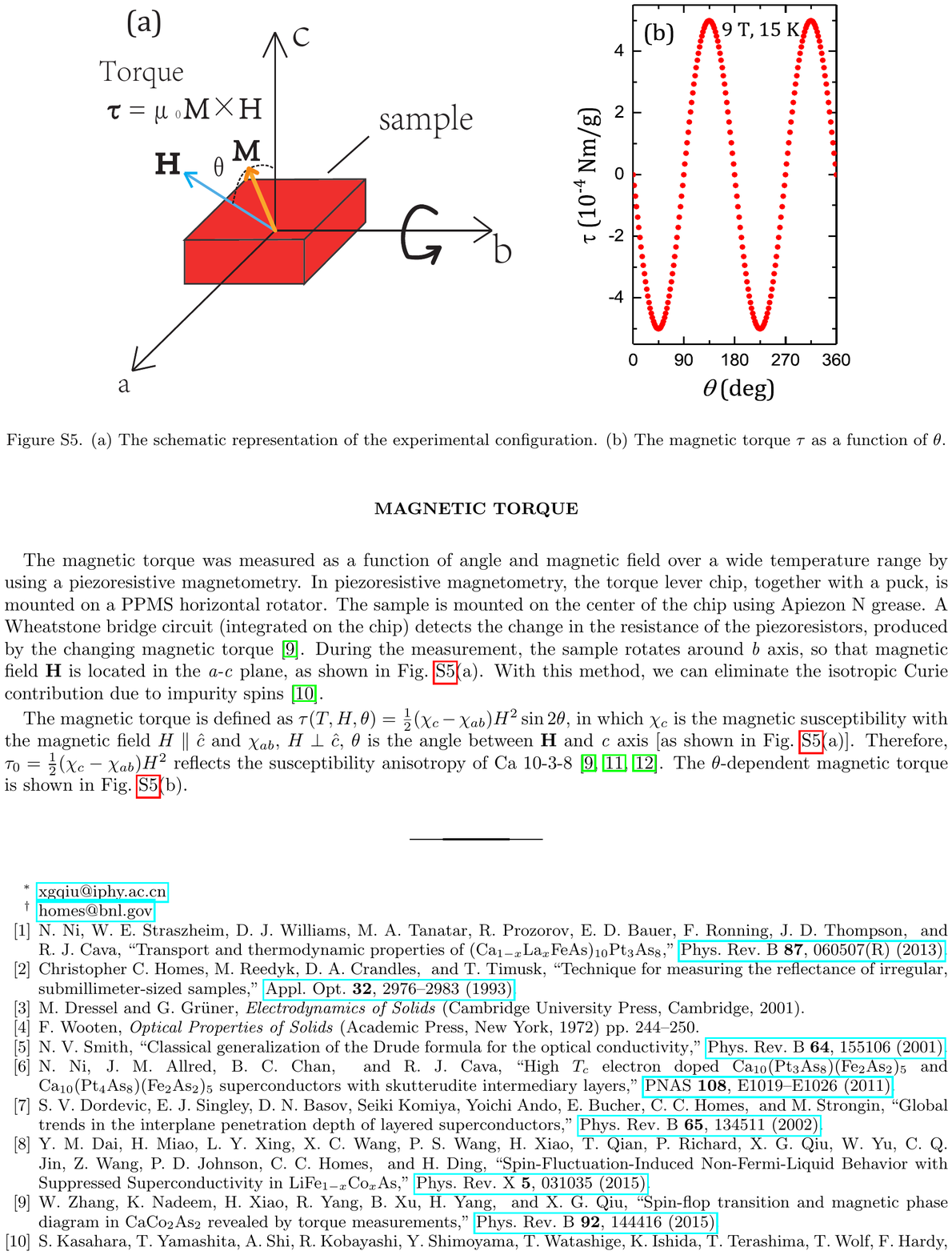} \\
  \ \\
}

\newpage
\vspace*{-2.0cm}
\hspace*{-2.5cm} {
  \centering
  \includegraphics[width=1.2\textwidth]{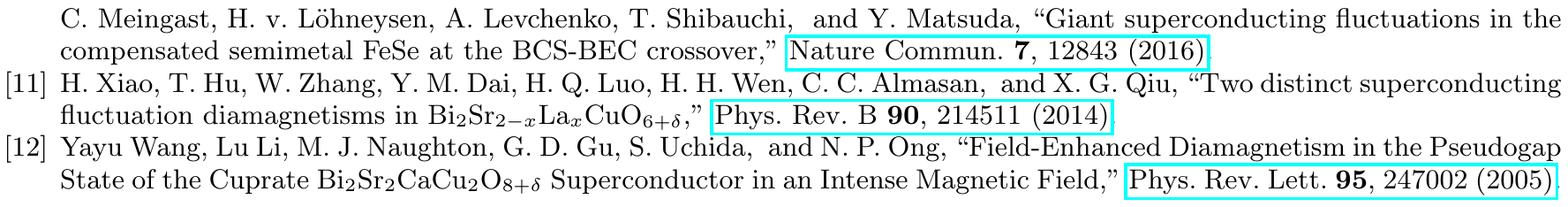} \\
  \ \\
}

\end{document}